\documentclass[aps,prl,twocolumn,showpacs]{revtex4}
\usepackage{graphicx}
\usepackage{dcolumn}
\usepackage{bm}

\begin{document}

\newcommand{\smco}{SmCo$_5$}
\newcommand{\w}{$\omega$}
\newcommand{\m}{$\mu$}
\newcommand{\sdr}{$\sigma \cdot r$}
\newcommand{\vdip}{$\mu$}

\title{Short-range, spin-dependent interactions of electrons: a probe for exotic pseudo-Goldstone bosons}

\author{W. A. Terrano}
\author{E. G. Adelberger}
\author{J. G. Lee}
\author{B. R. Heckel}

\affiliation{Center for Experimental Nuclear Physics and Astrophysics, Box 354290,
University of
Washington, Seattle, Washington 98195-4290}
\date{\today}

\begin{abstract}
We used a torsion pendulum and rotating attractor with 20-pole electron-spin distributions to probe dipole-dipole interactions mediated by exotic pseudo-Goldstone bosons with $m_{\rm b}\leq 500 \; \mu$eV 
and coupling strengths up to 14 orders of magnitude weaker than electromagnetism. This corresponds to
symmetry-breaking scales $F \leq 70$ TeV, the highest reached in any laboratory experiment. We used an  attractor with a 20-pole unpolarized mass distribution to improve laboratory bounds on $CP$-violating monopole-dipole interactions with $1.5\:\mu$eV$<m_{\rm b}<400\:\mu$eV by a factor of up to 1000.
\end{abstract}

\maketitle

Spontaneously-broken global symmetries
play an important role in particle physics\cite{we:72}. When the underlying symmetry is exact, the process always produces massless pseudoscalar Goldstone bosons whose coupling to a fermion with mass $m_f$ is 
$g_{\rm p}\!=\!m_{\rm f}/F$, where $F$ is the energy scale of the spontaneously broken symmetry.
If the symmetry is not exact but explicitly broken as well, as in the chiral symmetry of QCD, the fermionic couplings are unchanged, but the resulting pseudo-Goldstone bosons, such as the QCD pions, acquire a small mass $m_{\rm b}=\Lambda^2/F$ where $\Lambda$ is the explicit symmetry-breaking scale of the effective Lagrangian. Searches for the ultra-weak, long-range interactions mediated by exotic pseudo-Goldstone bosons, therefore, provide very sensitive and general probes for new hidden symmetries broken at extremely high energies. 

\begin{figure}[b]
\hfil\scalebox{.45}{\includegraphics*{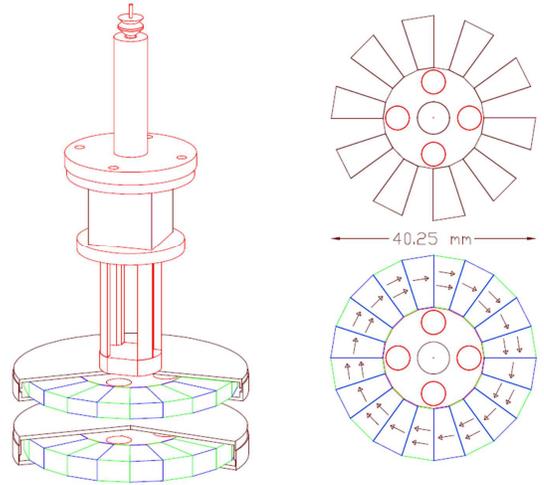}}\hfil
\caption {Left: the 20-pole spin pendulum and spin-attractor; $\mu$-metal cans on the pendulum and attractor are cut away to show the Alnico (green) and \smco\  (blue) segments and one of the 4 pairs of calibration cylinders (red). The mirror cube allowed us to monitor the pendulum twist angle. The magnetic shield surrounding the entire pendulum is not shown. 
 Lower and upper right: top views of the spin and mass attractors, respectively.  Arrows indicate net spin density and direction.}
\label{fig: pendattr}
\end{figure}

The tree-level potentials from pseudoscalar boson exchange 
are purely spin-dependent. The classic pseudoscalar potential is the dipole-dipole interaction 
\begin{eqnarray}
V_{\rm dd}&=&\frac{g_{\rm p}^2 \hbar^2}{16\pi m_e^2 c^2 r^3}\left[(\bm{\hat{\sigma}_1 \cdot \hat{\sigma}_2})
\left( 1+\frac{r}{\lambda}\right)\right.~~~~~ \nonumber \\ 
~~&-&3\left.(\bm{\hat{\sigma}_1 \cdot \hat{r}})(\bm{\hat{\sigma}_2 \cdot \hat{r}})
\left(1+\frac{r}{\lambda}+\frac{r^2}{ 3\lambda^2}\right) \right] 
e^{-r/\lambda}~,~~~~~
\label{eq: V1}
\end{eqnarray}
where $\lambda=\hbar/(m_{\rm b} c)$. Axion-like bosons with an additional scalar coupling, $g_{\rm S}$, can also generate a monopole-dipole interaction\cite{mo:84}
\begin{eqnarray}
V_{\rm md}  = \frac{\hbar g_{\rm s} g_{\rm p}}{8 \pi m_{e} c} \left[(\bm{\hat{\sigma} \cdot \hat{r}}) \left( \frac{1}{r \lambda } + \frac{1}{r^2} \right) \right]e^{-r/\lambda}~.
\label{eq: V2}
\end{eqnarray}
Because these potentials average to zero for unpolarized bodies, traditional searches for new macroscopic forces are essentially insensitive to such bosons. 
Motivated by theoretical conjectures that propose additional pseudo-Goldstone bosons such as  axions, familons, majorons, closed-string axions and accidental pseudo-Goldstone bosons (see \cite{ri:04} for a recent review), we developed a generic ``pseudo-Goldstone detector'' with high sensitivity to both $V_{\rm dd}$ and $V_{\rm md}$ interactions. 
We combined the strategies of previous E\"ot-Wash torsion-balance probes of electron-spin-dependent forces\cite{he:08,he:13} (closed magnetic circuits containing high and low spin-density materials) and short-distance gravity\cite{ho:04,ka:07} (planar geometry, 
high-multipolarity signals and continuously rotating attractors) to produce the torsion-pendulum/rotating-attractor instrument shown in Fig.~\ref{fig: pendattr}. The small scale of  our device 
allowed us to probe $V_{\rm dd}$ interactions with $m_{\rm b}\leq 500\;\mu$eV.
Previous studies with polarized electrons\cite{he:13} and neutrons\cite{va:09,gl:08} were restricted to $m_{\rm b} \lesssim 5\;\mu$eV.
%
%

The key element of our instrument was a spin-ring containing 20 
equally-magnetized segments of alternating high and low spin-density materials. This formed a spin 20-pole with a negligible external magnetic field; the sensitivity to 
$V_{\rm dd}$ or $V_{\rm md}$ interactions 
arises entirely from the spin-density contrast in the rings. One spin-ring was the active element of our detector, a torsion pendulum 
placed just above an attractor rotating at frequency $\omega$. Our dipole-dipole search used an attractor consisting of a second 20-pole spin-ring so that 
$V_{\rm dd}$ interactions would produce a $10\omega$ torque on the pendulum. 
We used an unpolarized copper attractor that formed a mass 20-pole to measure the gravitational background in our $V_{\rm dd}$ study and to probe $V_{\rm md}$ interactions. The $10\omega$ torques from $V_{\rm md}$ and from gravity were distinguishable because their turntable phases differ by 9 degrees. 
The pendulum and both attractors each contained four 4.8 mm-diameter cylinders (tungsten and vacuum for the spin and mass attractors, respectively) that provided a continuous $4\omega$ gravitational calibration signals. 

Alternating wedges of SmCo$_5$ and Alnico provided the spin contrast of the rings.  SmCo$_5$ has a substantial orbital contribution to its magnetic field\cite{he:08} while
Alnico's magnetism comes almost entirely from polarized electrons. \smco\ fully magnetized to 9.8\,kG contains  $\sim4.5 \cdot 10^{22}$~spins/cm$^3$ while Alnico magnetized to the same degree has $\sim 8 \cdot 10^{22}$~spins/cm$^3$ \cite{he:08}.  We cut the SmCo$_5$ wedges from commercially magnetized material while the Alnico was magnetized simply by assembling the ring.
We tuned the precise magnetization of each Alnico wedge {\em in situ} by applying a localized external field 
until the peak-to-peak leakage field 3\,mm from the ring 
was reduced from $\sim$100\,G to $\sim$8\,G.
We then enclosed the ring assemblies in nested 2-layer $\mu$-metal cans with a total thickness of 0.53\,mm, reducing the peak-to-peak residual field to $\sim 10\,\mu$G.
A 0.99\,mm thick shielding screen consisting of 21 layers of alternating $\mu$-metal and aluminum foils
separated the attractor assembly from the pendulum. A 1.27\,mm thick, cylindrical, $\mu$-metal ``house"  surrounded the pendulum except for a hole for the suspension fiber and another that provided optical access to the pendulum; a 0.76\,mm-thick $\mu$-metal tube surrounded the attractor turntable. 
The peak-to-peak magnetic field change at the pendulum location with the full shielding in place was below our resolution of 2\,\m G. 

Could the magnetic shielding also shield the $V_{\rm dd}$ and $V_{\rm md}$ interactions? We discuss this important point elsewhere\cite{he:15} and show that such a shielding effect is completely negligible in our case. 

The mass-density difference between \smco\ (8.31 g/cm$^3$) and Alnico (7.31 g/cm$^3$) would produce a significant gravitational $10\omega$ torque.
We placed 76\,$\mu$m thick W shims above and below each Alnico segment and Ti shims above and below each \smco\ segment
to minimize $10\omega$ gravitational torques.

The active elements of the pendulum and attractor, along with the magnetic shielding, were installed in a rotating-attractor torsion balance normally used to study short-distance gravity. Details of that torsion balance and the general methods of data analysis are given in Refs.\cite{ho:04,ka:07}. The only other significant change to the apparatus described in Ref.~\cite{ka:07} 
was an improved attractor drive that locked the output of a 
$2^{21}$ pulses/rev angle-encoder to a crystal-controlled oscillator\cite{co:13}. 

We centered and leveled the attractor ring to $\sim15\,\mu$m of the turntable rotation axis with optical and mechanical techniques.  We leveled the pendulum to 130$\,\mu$m of its rotation axis using capacitive techniques, and centered it to about $\pm 20\mu$m of the attractor rotation axis by maximizing the 4$\omega$ signal from the mass attractor.
These misalignments were negligible relative to our typical separations $s\gtrsim 2$ mm. 
We inferred the vertical separation, $s$, between the bottom of the pendulum ring magnets and the top of the attractor magnets (or copper) using a $z$-micrometer on the vertical translation stage that supported the suspension fiber. We
measured the attractor-pendulum capacitance as a function of $z$ and fitted these data to a finite-element electrostatic model to map the $z$-micrometer readings into the pendulum-screen separation. Mechanical and optical measurements provided the additional information needed to determine $s$.  

The torque on the pendulum was inferred from a harmonic analysis of its twist angle, $\theta$, as a function of the attractor angle $\phi=\omega t$. 
The data analysis procedure was similar to that used in Ref. \cite{ka:07}. 
The $\theta(\phi)$ time series was processed by a 4-point digital filter that suppressed free torsional oscillations as well as the DC response and linear drift. 
Data runs were divided into cuts containing exactly one attractor revolution and 
each cut was fitted with a quadratic drift term plus the first 14 harmonics of the turntable angle. The harmonic amplitudes were corrected for pendulum inertia, electronic time constants and the response of the digital filter, and then converted into torques using the effective value of the fiber's torsional constant $\kappa=I(2\pi/T_0)^2=3.1$ aN\,m/nrad,
where the $T_0$ is pendulum's free oscillation period and its moment of inertia, $I\!=\!134~{\rm g}\,$cm$^2$, was computed from a detailed numerical model. $T_0$ was determined from ``sweep runs'' taken after each science run; the attractor turntable was stopped and the pendulum was given a $\sim 10 \mu$rad kick. The resulting  oscillations were analyzed to obtain a precise period and, crucially, to map out small nonlinearities in the autocollimator's analog position-sensitive detector. The measured 10$\omega$ and 4$\omega$ torques from a run were found by weighting equally all (typically 48) cuts in that run with the statistical uncertainties determined by the scatter of the results. These data were compared to those expected from $V_{\rm dd}$ and $V_{\rm md}$ interactions and from gravity. The expected torques, assuming that the pendulum was aligned with the attractor, were computed using the Fourier-Bessel expansion which converges rapidly for our application.

\begin{figure}[t]
\hfil\scalebox{.46}{\includegraphics*{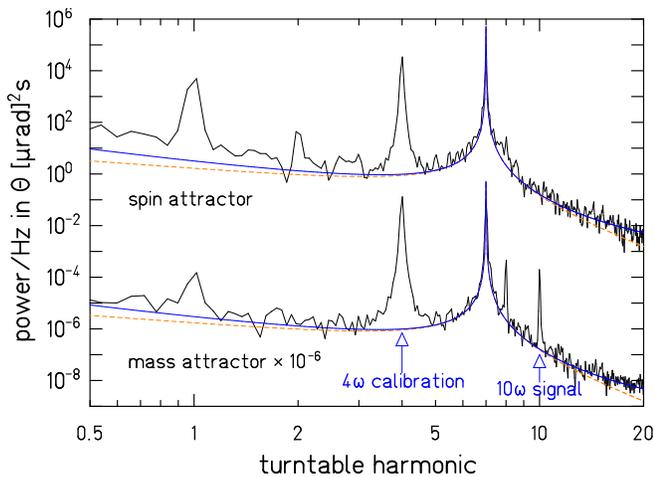}}\hfil
\caption{Sample power spectral densities of the twist signals from the spin and mass attractors at the closest attained separations. The dashed lines show the thermal noise, solid lines include the effect of an additional $1/f^2$ component. The 8\w~peak in the mass attractor data is the 1st harmonic of the calibration signal. It is much smaller in the spin attractor data because of its larger value of $s$.}
\label{fig: psd}
\end{figure}

The attractor rotation periods, $T_{\rm att}=7T_0$ or $6T_0$, were selected to place our $4\omega$ and $10\omega$ signals in low-noise regions (see Fig. \ref{fig: psd}). 
The noise was dominated by thermal fluctuations from internal losses in the suspension fiber which gave the torsion oscillator a quality factor of
$Q\approx1500$. 

%
%
%
%
\begin{table}[b]
\caption{Observed $4\omega$ and $10\omega$ torques. Amplitudes $A$, are in units of  aN\,m, phases $\phi$ are in degrees, and separations $s$ are in mm. The $1\sigma$ uncertainties do not include systematic effects. If $V_{\rm md}\!=\!0$, we expect $\Delta\phi=
\phi_{10\omega}\!-\!\phi_{4\omega}\!=\!-9.0^{\circ}$.}
\begin{ruledtabular}
\begin{tabular}{ccccc}
attractor   & $n$    & $A_{4\omega}$        & $A_{10\omega}$ &  $\phi_{10\omega}-\phi_{4\omega}$ \\
\colrule
spin:~~$s\!=\!4.12$  & 7    &  $\!2855\!\pm\!5$ & $0.7\!\pm\! 2.9$  &  $+3\!\pm\!25$ \\
spin:~~$s\!=\!4.12$  & 6    &  $\!2863\!\pm\!4$ & $2.9\!\pm\! 2.8$  &  $-7.9\!\pm\!5.5$ \\
spin:~~$s\!=\!4.12$  & 7+6  &  $\!2860\!\pm\!3$ & $1.3\!\pm\! 2.0$  &  $-6.1\!\pm\!8.6$ \\
mass:~~$s\!=\!1.98$  & 7    &  $\!5611\!\pm\!8$ & $344\!\pm\!4$  &  $-9.47\!\pm\!0.08$ \\
\end{tabular}
\end{ruledtabular}
\label{tab: torques}
\end{table}
%
%
%
\begin{figure}[t]
\hfil\scalebox{.4}{\includegraphics*{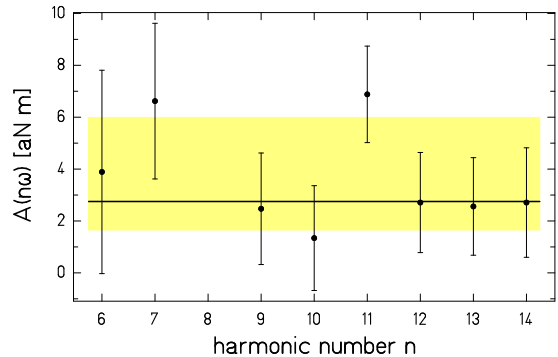}}\hfil 
\caption{Comparison of the spin-attractor $10\omega$ science signal with nearby background signals. The shaded horizontal band indicates the mean and standard deviation, $\sigma$, of the background signals. The horizontal line shows the mean amplitude, $\sigma\sqrt{\pi/2}$, expected for random signals whose quadrature components have zero mean and spread $\sigma$.}  
\label{fig:mag}
\end{figure}

\begin{figure}[t]
\hfil\scalebox{.56}{\includegraphics*{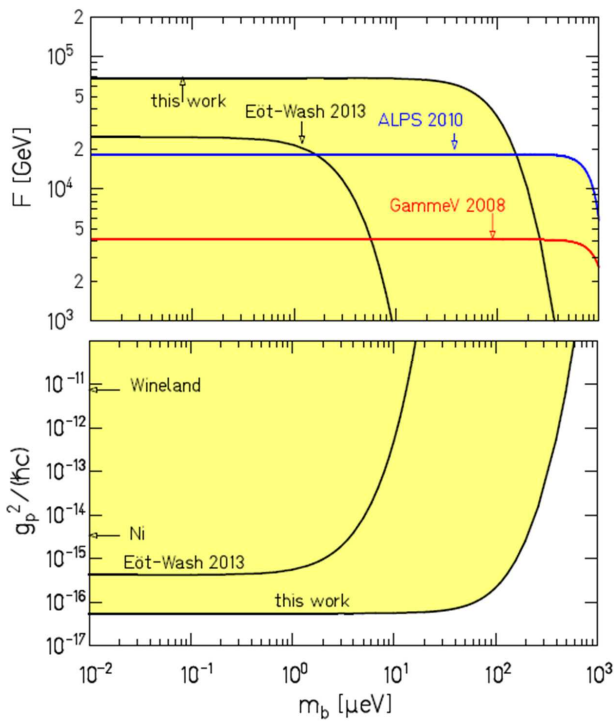}}\hfill
\caption{Bottom: exotic dipole-dipole limits from this work and Ref. \cite{he:13}. Arrows indicate the infinite-range constraints from Refs.~\cite{wi:91,ni:94}. Electron $g\!-\!2$ constraints are at the $10^{-10}$ level\cite{gminus2}. Top: limits on the symmetry-breaking scale from this work and Refs. \cite{ch:08,eh:10}.  The shaded areas are excluded at $2\sigma$.}
\label{fig: dd limits}
\end{figure}

Spin-attractor data were taken at $s=4.12$, 5.13, and 8.15 mm (uncertainties are $\pm 0.015$ mm).  Because of the $1/r^3$ fall off of the potential, our $V_{\rm dd}$ sensitivity comes entirely from the $s=4.12$~mm data. The results from 165 hours of $s=4.12$ mm data are shown in Table~\ref{tab: torques}. 
We expected the largest systematic effects with the spin attractor to be residual gravitational and magnetic couplings betweens the pendulum and the attractor.  Mass attractor data supplemented by calculations showed that our shims reduced the gravitational component of A$_{10\omega}$ by two orders of magnitude to $\sim$1\,aN\,m.  Measurements showed that the magnetic leakage field was fairly constant across all higher ($>$ 5\w ) harmonics. We observed little evidence for such couplings (see Fig.~\ref{fig:mag}), which would have produced torques
at all these frequencies as well as at 10\w. As a result, no corrections
for magnetic backgrounds were necessary.
Other systematic concerns such as thermal and electrostatic effects were found to be negligible.
Our final value is $A_{10\omega} = (1.3\pm 2.2)$~aN\,m for a 2$\sigma$ upper limit of 5.1\,aN\,m. 
The corresponding constraints on a new dipole-dipole interaction, and the associated bounds on the
symmetry-breaking scale $F$, are shown in Fig \ref{fig: dd limits}. 
%
%
These are the most sensitive laboratory constraints on $(g_{\rm p}^e)^2/\hbar c$ 
for $m_{\rm b}\leq 500\:\mu$eV/c$^2$ (at the $5.5 \times 10^{-17}$ level for $m_{\rm b} < 30\; \mu$eV). To our knowledge, the only other laboratory constraints on pseudoscalars in this mass range are Ramsey's 1979 limit, $(g_{\rm p}^p)^2/\hbar c <3 \times 10^{-4}$ level\cite{ra:79}, on anomalous spin-spin interactions between protons. Our results indicate that $F>70$ TeV.

Because magnetic backgrounds in our $V_{\rm md}$ study using the mass attractor were small, we could use a single 0.25 mm thick \m -metal screen. This allowed us to take mass-attractor data at $s=1.98$ mm as well as at 2.03, 3.00, 4.04 and 7.99 mm (uncertainties are $\pm .015$ mm). Our V$_{\rm md}$ constraints come entirely from the $s=1.98$\,mm data.  The other data allowed us to check for systematics and validate our gravitational calculations.  The observed $4\omega$ and $10\omega$ torques from 38 hours of $s=1.98$ mm data are shown in Table~\ref{tab: torques}.  The gravitational contribution to the $10\omega$ torque vanishes when the copper arms of the attractor are directly below either the \smco\ or the Alnico. This occurs (see Fig.~\ref{fig: pendattr}) when the attractor is rotated 
$9^{\circ}$ away from the angle at which calibration cylinders are aligned. Conversely, the $V_{\rm md}$ torque is maximal at those orientations. This allowed us to separate the effects of gravity from a $V_{\rm md}$ interaction. The $V_{\rm md}$ component of the $10\omega$ torque is 
\begin{equation}
A_{\rm md}=A_{10\omega}|\sin 10(\Delta \phi+\delta\phi)|~,
\end{equation} 
where $\Delta\phi= \phi_{10\omega}-\phi_{4\omega}$ and $\delta\phi$ is nominally $9^{\circ}$. 
The absolute value occurs because of the 4-fold ambiguity in the attractor angle inferred from the $4\omega$ signal.
Our $A_{\rm md}$ bound is dominated by the systematic uncertainty in $\delta\phi$.
Alignment microscope measurements showed that phase of the magnet ring relative to the calibration cylinders was only fixed to $\pm 0.17^\circ$. 
%
%
\begin{figure}[b]
\hfil\scalebox{.58}{\includegraphics*{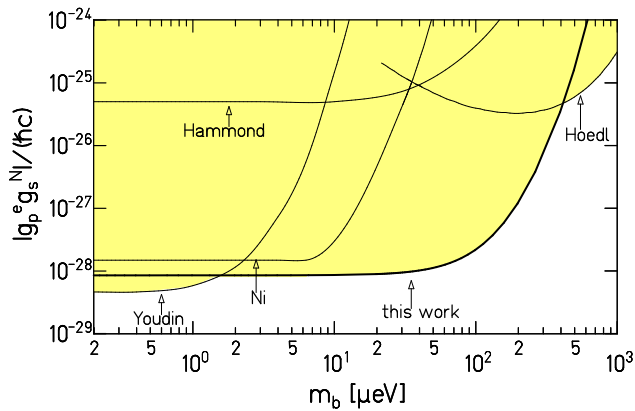}}\hfill
\caption{Monopole-dipole constraints from this work and refs.\cite{yo:96,ni:99,ha:07,ho:11}  The shaded region is excluded at $2\sigma$. The $m_{\rm b}=0$ limit from Ref.~\cite{he:08}is $2\times 10^{-36}$. (We doubled the $1\sigma$ limits given in refs.\cite{yo:96,ho:11}.)}
\label{fig: md limits}
\end{figure}
An estimated $50\,\mu$m accuracy in positioning the gravitational shims, revealed by the behavior of $\phi_{10\omega}$ in our centering data, contributed an additional error of $\pm 0.29^\circ$ and increased the uncertainty in $\delta\phi$ to $\pm 0.34^\circ$. 
This gives a $1\sigma$ result 
$A_{\rm md} = (18 \pm 12)$~aN\,m with a 2$\sigma$ limit, $A_{\rm md} \leq 38$\,aN\,m. Our $|(g_{\rm p}^e g_{\rm s}^N)|/\hbar c$ constraint, shown in Fig. \ref{fig: md limits}, improves upon previous work by up to a factor of 1000 for $1.5\leq m_{\rm b}\leq 400$ \m eV/c$^2$. The most sensitive limit on $(g_{\rm p}^n g_{\rm s}^N)/\hbar c$ is also at the $10^{-28}$ level\cite{tu:13}.
 
Stellar cooling rates\cite{ra:95} constrain $V_{\rm dd}$ interactions of simple pseudoscalar particles at a level well below our bound, and 
 the astrophysics bound on $g_{\rm p}^e$, combined with bounds
on $g_{\rm s}^N$ from gravitational experiments, set very tight limits on $V_{\rm md}$ interactions between electrons and nucleons\cite{ra:12}. However, 
a chameleon mechanism could invalidate these astrophysical bounds while having a negligible effect in cooler, less dense lab environments\cite{ja:06}. In this case $V_{\rm dd}$ and $V_{\rm md}$ can only be constrained by laboratory experiments such as this work which reveals that any hidden symmetry involving electrons must be broken at an energy scale $F > 70$ TeV and, if it is explicitly broken as well, that scale $\Lambda$ must be  $>0.1$ MeV.  These set the highest laboratory bounds on the minimum energy scale of new hidden symmetries involving leptons. Extensions of general relativity that include torsion as well as  curvature  predict infinite-range dipole-dipole interactions\cite{ni:10} and are also constrained by this work.

We thank T.S. Cook, S.M. Fleischer and H.E. Swanson for 
assistance with the apparatus, C.A. Hagedorn and Amol Upadhye for helpful conversations. This work was supported by NSF Grants PHY0969199 and PHY1305726 and by the DOE Office of Science.

\end{document}